\documentclass{aa_2012}

\usepackage{natbib}
\bibpunct{(}{)}{;}{a}{}{,} 

\usepackage{xcolor}
\usepackage{txfonts}
\usepackage{graphicx}
\usepackage{epsfig}
\usepackage{natbib}

\usepackage{url}
\usepackage{twoopt}
\usepackage{dcolumn}
\usepackage[colorlinks=true]{hyperref}
\hypersetup{citecolor=blue}

\begin{document} 

\title{Line shapes of the Na/K resonance line profiles  perturbed by
  H$_2$ at extreme density
 \thanks{Complete K-H$_2$ opacity tables for the D1 and D2 components of the
       resonance line are only available at the CDS
    via anonymous ftp to cdsarc.u-strasbg.fr (130.79.128.5) }}

\subtitle{}

\author{N. F. Allard         \inst{1,2}
   \and J. F. Kielkopf        \inst{3}  
}     

\institute{LIRA, Observatoire de Paris, Universit\'e PSL, Sorbonne
     Universit\'e, Sorbonne Paris Cit\'e, CNRS,
     61, Avenue de l’Observatoire, F-75014 Paris, France\\
              \email{nicole.allard@obspm.fr}
         \and
         Institut d'Astrophysique de Paris,  UMR7095, CNRS, 
         Universit\'e Paris VI, 98bis Boulevard Arago, F-75014 PARIS, France \\
          \and
          Department of Physics and Astronomy, 
          University of Louisville, Louisville, Kentucky 40292 USA \\       
}

   \date{Received 21 August/ Accepted 1 October 2025}

   \abstract
{Collision broadening by molecular hydrogen of  sodium and potassium 
  is one of the major broadening mechanisms in the atmospheres of brown dwarf
  stars and exoplanets
  at an effective temperature of about  1000~$\ \mathrm{K}$.
The relevant H$_2$ perturber densities reach several $10^{19}$~cm$^{-3}$
  in hot ($T_{\rm eff}\gtrsim 1500$ K) Jupiter-mass planets, and up to almost 
  $10^{21}$cm$^{-3}$ ($\approx 30$~bar) for  more massive or cooler objects.
The Juno Microwave Radiometer has enabled observations of
Jupiter’s atmosphere down  to previously inaccessible depths where pressures near $10^6$~bar  have to be considered and the relevant
H$_2$ perturber densities may exceed $10^{25}$ cm$^{-3}$.
}
{While Na/K-He/H$_2$ opacity tables have been constructed for the resonance lines
that are valid to $n_{\mathrm{H}_2}=10^{21}$~cm$^{-3}$,
at higher density it  is important
to ensure accurate absorption cross-sections of these species in the models.
We accurately determine the broadening of Na/K  by H$_2$ in
the  unified theory  at  H$_2$ densities larger
than $10^{21}$~cm$^{-3}$ and compare to the corresponding Lorentzian profiles.
}
{The theory of spectral line shapes, especially the unified approach we have
developed, makes possible accurate models of stellar spectra that account both
for the centers of spectral lines and their extreme wings in one consistent
treatment. In this study, we examine the density dependence  of
the Na and K $D$2 ($P_{3/2}$) components, respectively, at
5889.95~\AA\/ and 
7664.90~\AA\/  from $1 \times 10^{21}$ to $2 \times 10^{22}$~cm$^{-3}$.
 }
{Lorentzian profiles from impact  broadening
theory are only valid in the core of the line not farther than a
few half-widths as long as there is no overlap between the core of the
line and possible quasi-molecular features in the wings due to close
collisions. 
}
{The accurate computation of line profiles from collision broadening at high density requires use of a Fourier transform of the autocorrelation function inside
  the model atmosphere code. We strongly  warn that use of Lorentzian profiles at a high perturber density neglects radiation during close collisions and may lead to erroneous conclusions.
}

\keywords{Lines: profiles brown dwarfs, planets
  planets: atmospheres.}

\authorrunning{N.~F.~Allard and J.~F. Kielkopf}

 \titlerunning{Na/K resonance line profiles  perturbed by high
     pressure of H$_2$}
 
   \maketitle

 \section{Introduction}
   
An understanding of how ambient molecules, atoms, ions, and electrons affect spectra of absorbing or emitting atoms is invaluable within many fields of science. Often termed pressure broadening, its accurate modeling with fundamental physics is possible with validation through  laboratory experiments. It is often assumed that atomic spectral lines have Lorentzian profiles with width and shift proportional to the 
gas density. When that is the case, the resultant spectra are described by Voigt profiles for discrete transitions, and incorporation in stellar spectra models is computationally straightforward.  However, we know from the underpinning physics that many spectral lines exhibit asymmetric profiles and diffuse ``satellite'' bands associated with atomic and molecular collisions that occur during the radiative process. While satellite bands were first reported in the late 1920's,  their explanation evolved with improvements to the theory of spectral line formation and its dependence on the binary interactions of the emitter with its transient perturbers. The field  has been summarized in a number of review articles \citep[e.g.,][]{margenau1936,chen1957,allard1982}. Given curated atomic and molecular data, astrophysicists use the results of these experimental and theoretical studies to aid them in determining the temperature, pressure and composition of planetary and stellar atmospheres. An example from our prior work is the appearance of Lyman~$\alpha$ line satellites in synthetic spectra of DA white dwarf stars \citep[e.g.,][]{allard1992}. The purpose of this brief paper is to assist stellar and planetary astrophysicists with the implementation of state of the art line shape calculations in these applications.

As an illustration, consider why the unique optical spectra of L and T-type
low-mass and brown dwarf stars exhibit a depressed continuum from the visible
through the near-infrared dominated by the far wings of the absorption
profiles of the Na 3$s$-3$p$ and K 4$s$-4$p$ doublets perturbed by
molecular hydrogen and helium.
 The $P_{1/2}$ ($D1$) line is due to a simple isolated $A$ $\Pi_{1/2}$ state,
the line profile for the $D1$ line is totally asymmetric,
whereas the $P_{3/2}$ ($D2$) line comes from the $A$ $\Pi_{3/2}$ and 
$B$ $\Sigma_{1/2}$ adiabatic states arising from the $4p$ $P_{3/2}$ atomic 
state. Blue satellite bands in alkali-He/H$_2$ profiles
are correlated with maxima in the excited $B$ $\Sigma_{1/2}$
state potentials and can be predicted from the  maxima in 
the difference potentials $\Delta V$ for the $B$-$X$ transition \citep{kielkopf2017}.
Our Na/K--He/H$_2$ opacity tables of absorption coefficients were first
implemented in \citet{allard2003} in stellar model atmospheres  and
synthetic spectra by  using the  \citet{allard2001} PHOENIX code.
The results, compared to prior work lacking this fundamental spectral line theory, improved on quantitative interpretation of the observed
spectra \citep{allard2003,allard2007b}. 
Subsequently, \citet{tinetti2007} presented the first applications of 
profiles of Na and K perturbed by H$_2$ to the modeling of structure
and spectra of extra-solar planet atmospheres.
Furthermore, it is known that the interiors of giant planets
may contain deep radiative zones \citep{guillot1994,siebenaler2025} that
influence the transport of
material from the interior to the atmosphere,
requiring opacities valid at high density to model correctly.
The deep structure that determines the internal processes and external
emergent spectra may be evident in data from the Juno Microwave Radiometer
that was designed to see deeper into Jupiter’s atmosphere than any
previous instrument \citep{janssen2017,aglyamov2025}.
Theoretical models of the interiors of giant planets depend on the
properties of H$_2$/He at pressures approaching  $10^6$~bar and temperatures
of 5000 K \citep{stevenson1982,militzer2024}.
For these applications to stellar and giant planet atmospheres,
opacity tables  were constructed allowing line profile models up
to $n_{\mathrm{H}_2} = 10^{21}$ cm$^{-3}$ over a wide range of temperatures
to be obtained.  It is at the highest densities that multiple-perturber
effects have to be included because the line shapes are extremely
non-Lorentzian.  However, the tables also provide precise representation
of the broadening, shift and developing asymmetry from the lowest densities
contributing across the optical and infrared spectrum.

Similarly,  \citet{homeier2007} investigated these effects on the Na doublet observed in the spectra of 
metal-rich white dwarfs with a He-dominated atmosphere using opacities calculated from fundamental Na-He neutral atom interactions as input to spectral line shape theory. They found that cool white dwarfs show very strong
Na absorption developing where the atomic helium density could reach beyond 10$^{21}$ to 10$^{22}$~cm$^{-3}$.
High helium density effects on the Na resonance line were  presented in
Fig.~7 of \citet{allard2023} for He pressure up to 400 bar. 
Accurate profiles that are valid at such 
high densities of helium are required to  be incorporated into white dwarf spectral
models \citep{blouin2019c}.

The opacity in these cases cannot be represented analytically with a simple Lorentzian, Gaussian, or Voigt profile.  Such often-used approaches are valid at low density if the width, shift and possible asymmetry parameters are calculated by accurate atomic physics, but the caveat is that is rarely done.  Precise parameters are known only for limited pairs of atomic species such that a model with millions of atomic lines emitted by the whole range of atomic abundances, temperatures, and electron densities depends on estimates or ad hoc assumptions that are not comprehensively validated.  When Doppler broadening due to high temperature, stellar rotation, and convection dominate the disk-integrated emergent spectrum, in practice this shortcoming can be inconsequential.  However, in a high density environment where the line strength of an abundant species is in a few spectral lines, the regions far from the line center develop as inner regions saturate.  This leads to an opacity that can  only be obtained by comprehensive line shape theory that itself depends on accurately determined atomic interactions.  
In our work, a unified line shape theory and a set of atomic interaction
potential energies model the entire line profile from the
impact-broadened line center to the far wing by using the Fourier
transform (FT) of the spectral line shape 
autocorrelation function. This formalism, for which complete details and a derivation are given in \citet{allard1999}, encompasses a classical treatment of statistics and radiation with a quantum treatment of atomic structure and interactions.   A briefer explanation of the theory was given in a recent paper \citep{allard2023}. For the purposes of spectral synthesis for astrophysical applications it is necessary to understand in simple terms why this is necessary, how to properly implement the opacities from a mesh of tabular data when needed, and the shortcomings of using Lorentzian line shapes lacking validated parameters.

\section{High density effects on the Na/K-H$_2$ line profile}

\subsection{Spectral line absorption}
\label{sec:prof}

There are various theoretical approaches for treating the problem of atomic lines broadened by collisions with atomic or molecular perturbers.
Fully quantum-mechanical methods  have been developed that 
allow practical calculations within the one-perturber approximation 
\citep{julienne1986,mies1986,herman1978,sando1983}.
The one-perturber approximation neglects the contribution of the core of
the line, and  is only valid   in the wing and in the limit of
densities low enough such that multiple-perturber effects  may be neglected.
An exact methodology for the quantum calculation at high densities,
as encountered in stellar atmospheres or laboratory plasmas, is not known. 
Approximate unified-theory methods to construct density-dependent line shapes
from one perturber spectra were developed in work by 
\citet{fano1963}, \citet{jablonski1945,jablonski1948},
\citet{anderson1952}, 
\citet{baranger1958a,baranger1958b,baranger1958c}, and \citet{szudy1975}.

Our theoretical approach is based on the quantum theory of
spectral line shapes of~\citet{baranger1958a,baranger1958b} 
in an  adiabatic representation to include the degeneracy of atomic
levels \citep{royer1974,royer1980,allard1994}. 
To evaluate an atomic spectral line shape, we computed a Fourier
transform written formally as
\begin{equation}
I(\Delta\omega)=
\frac{1}{\pi} \, Re \, \int^{+\infty}_0\Phi(s)e^{-i \Delta\omega s} ds,
\label{eq:int}
\end{equation}
where $s$ is time. The FT in Eq.~(\ref{eq:int}) 
is taken such that $I(\Delta\omega)$ is
normalized to unity when integrated over all frequencies,
and $\Delta\omega$ is the angular frequency difference measured relative to  the unperturbed spectral line. 

Here the correlation function $\Phi$ is not for a single isolated atom, but for
an ensemble of sources each experiencing a different microscopic environment
with a different temporal history during the radiative process.
The well-developed theory of spectral line shapes allows us to compute
the function
\begin{equation}
\Phi(s) = e^{-n_{p}g(s)},
\label{eq:phi}
\end{equation}
in which the density of perturbers $n_p$ is expressed explicitly, and the function
$g(s)$ depends only on single collisions.

The decay of the autocorrelation function $\Phi(s)$ with time
 leads to atomic line
broadening. It depends on the density of perturbing atoms $n_p$ and on their
interaction with the radiating atom.  Fundamentally we are able to calculate
a spectrum in this way for any neutral atom given the temperature, density, and
composition of the gas in which it is found. 
Moreover, the impact approximation determines the asymptotic 
behavior of the  unified line shape correlation function.  In this way
the results described here are applicable to a more general
line profile and opacity evaluation for the same perturbers 
at any given layer in the photosphere.
This approach to calculating the spectral line profile
requires knowledge of molecular 
potentials; that is, the binary interaction between an alkali atom and 
a perturbing atom as a function of their separation.   We use the
ab initio molecular-structure calculations  reported in 
\citet{allard2019} for the molecular potentials of
Na--H$_2$ system , and 
by \citet{allard2016b} for the K--H$_2$ system.

When $n_p$ is high, the spectrum is evaluated by computing the
FT of Eq.~(\ref{eq:phi}). The real part of $n_pg(s)$ damps
$\Phi(s)$ for large $s$ but this calculation is not  feasible when
extended wings have to be computed at low density  because of
the very slow decrease in the autocorrelation function.
In the planetary and brown dwarf upper atmospheres 
the H$_2$ density is of the order of $10^{16}$ cm$^{-3}$
in the region of line core formation but the far wings  play a crucial
role for the continuum generated far from the line center. 
An alternative is to use the expansion of the spectrum  $I(\Delta \omega)$
in powers of the density described in \citet{royer1971a}.
 For the  implementation of alkali lines perturbed by helium and molecular
hydrogen in atmosphere codes, the line opacity is calculated by splitting the
profile into a core component  described  with a Lorentzian profile, and
the line wings computed using an expansion of the autocorrelation function
in powers of density. 
This use of the density expansion is described in Section 3.3 of
\citet{allard2019}.
Opacity tables  of Na/K-He/H$_2$ that are the basis of line profiles
for the $D1$ and $D2$ components are   archived at the CDS
\citep{allard2016b,allard2019,allard2023,allard2024}.

\begin{figure}
 \centering
\resizebox{0.46\textwidth}{!}
{\includegraphics*{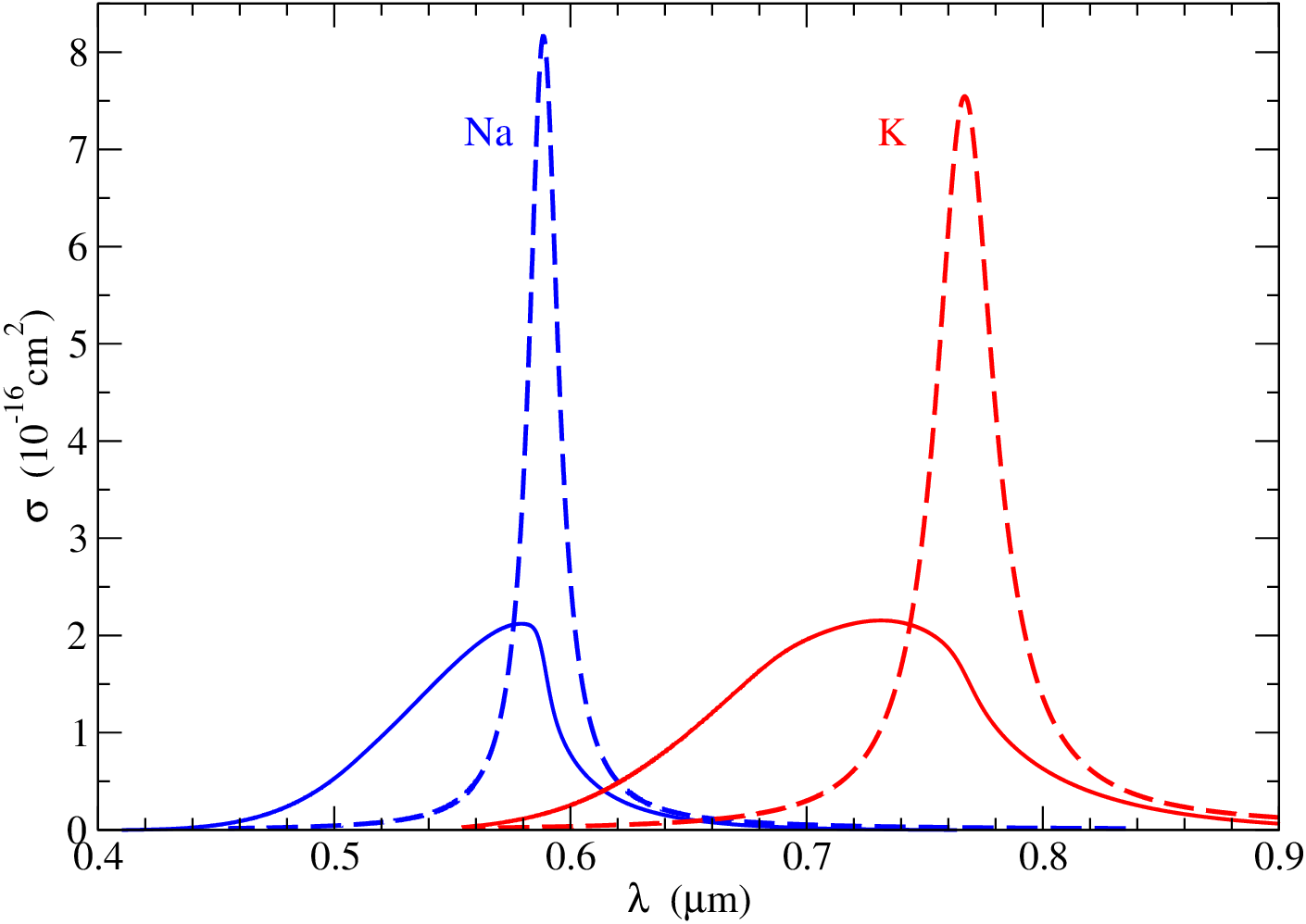}}
\caption  {Absorption cross section, $\sigma$,
  of  the $D$2  component of the  resonance lines of Na (blue curve) and
  K (red curve) perturbed by H$_2$ collisions at
$T = 1000$~K and $n_{\mathrm{H}_2} = 10^{22}$~cm$^{-3}$.
The Lorentzian approximation is overplotted for comparison (dashed curves).}
\label{fig:NaKH2lam}
\end{figure}

As a benchmark, the number density of a gas at 273.15~K and 100 kPa is Loschmidt's constant, exactly $2.651\,645\,804\times10^{19}$ cm$^{-3}$ \citep{codata2025}. An atmosphere of pressure is 101.325 kPa, referred to as 1 bar.  In laboratory experiments the density is increased largely by increasing pressure at approximately constant temperature, and consequently the increase in line widths is often referred to as pressure broadening. Line profile dependence on temperature enters through the relative kinetic energy of the perturbing atoms and the time dependence of binary collisions.  The profile's dependence on density results from the  collision  probability and, especially  at higher densities, the increased likelihood of several perturbing atoms acting simultaneously. We see that at densities well below  $10^{19}$ cm$^{-3}$ typical line profiles in the core of the line are Lorentzian, but as the density rises above $10^{21}$ cm$^{-3}$ the profiles become markedly non-Lorentzian, asymmetric, and may develop new components or ``satellites'' \citep{kielkopf1979,kielkopf1983}.    
Figure~\ref{fig:NaKH2lam} shows H$_2$ broadening   of the
$D2$ components of the resonance lines of Na and K
for a H$_2$ density $n_{H_2} =10^{22}$~cm$^{-3}$ (370~bar)
at 1000~$\ \mathrm{K}$.  

 The cross section shown in Fig.~\ref{fig:NaKH2lam} is determined from Eq.~\ref{eq:int} and the oscillator strength of the transition. The normalization of the line shape calculated with a FT given in Eq.~\ref{eq:int} connects to the absorption transition probability with the oscillator strength $f_{ij}$ for the transition between two states of the emitter-perturber system as given in CGS units by \citep{hubeny2015} (Eq. 5.105) for the total line absorption cross section
\begin{eqnarray}
\sigma_{ij} &=& \frac{h \nu_{ij}}{4 \pi}\,B_{ij}, \\
\sigma_{ij} &=& f_{ij} \frac{\pi e^2}{mc}, \\
\sigma_{ij} &=& \pi\,r_0\,f_{ij}.
\end{eqnarray}
$B_{ij}$ is the Einstein coefficient for photon absorption. In the last equation we have the convenient substitution of the classical radius of the electron $r_0 = 2.818 \times 10^{-13}$~cm.
The frequency-integrated opacity per distance through an absorbing atomic number density $n_a$  is then $\kappa = n_a \sigma$. 
The interpretation of this asymmetrical shape requires
us to consider the conditions for the distinctive appearance of  quasi-molecular features that are a consequence of radiative events during a close atomic collision.

\begin{figure}
\centering
\vspace{8mm}
\includegraphics[width=8cm]{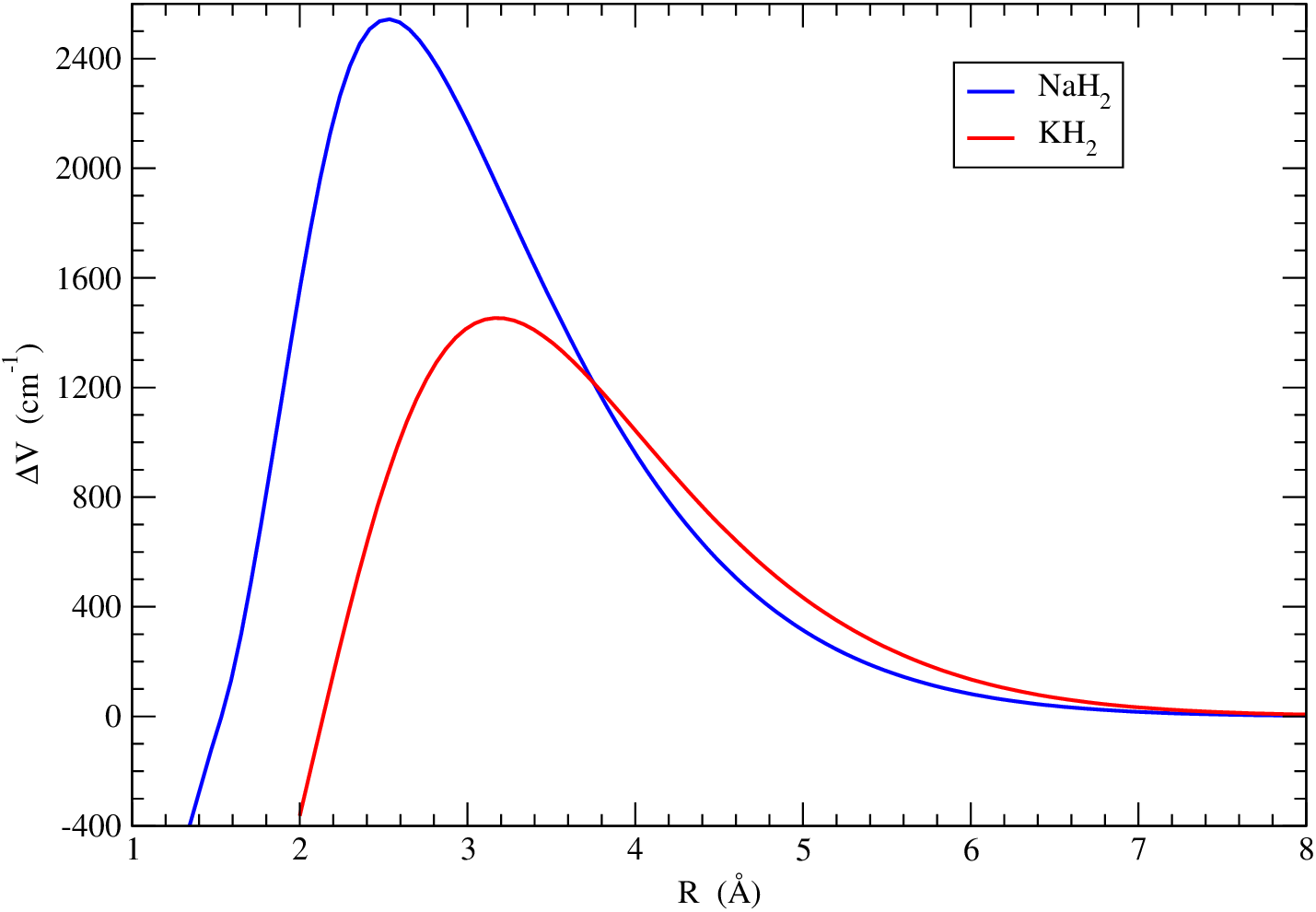}
\caption{$\Delta V(R)$  for the transitions  involved in the
  Na-H$_2$  line satellite (blue curve) and K-H$_2$  satellite (red curve).}
\label{fig:potdiff}
\end{figure}

\begin{figure}
 \centering
\resizebox{0.46\textwidth}{!}
{\includegraphics*{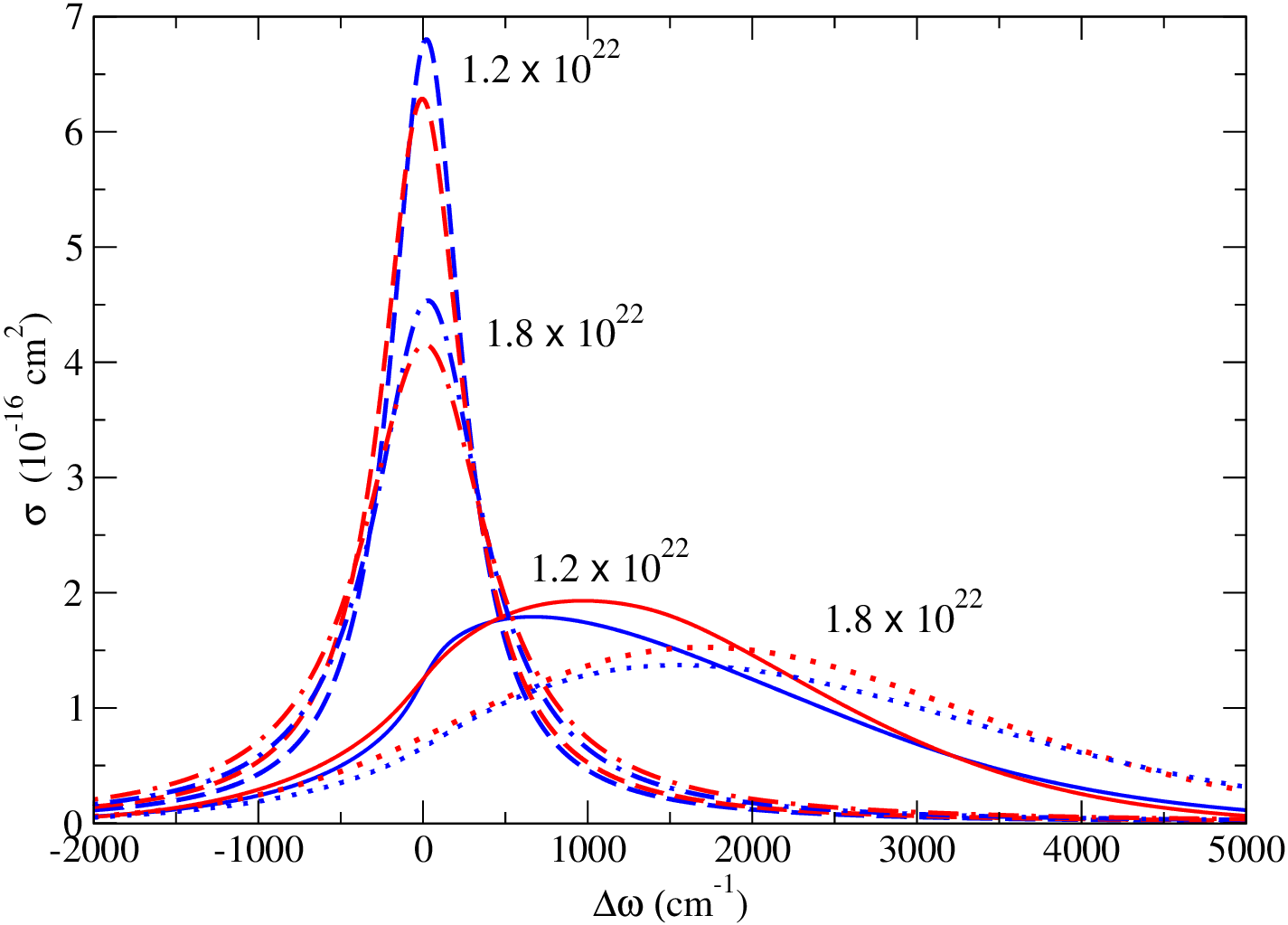}}
\caption  {Absorption cross section, $\sigma$, 
of the  $D$2  component of the  resonance lines of Na (blue curves )
and K (red curves) perturbed by H$_2$ collisions at 
$T = 1000$~K for $n_{\mathrm{H}_2} = 1.2 \times 10^{22}$ (full lines) and
$ 1.8 \times 10^{22}$~cm$^{-3}$ (dotted lines).
The Lorentzian approximation is overplotted for comparison (dashed curves).
$\Delta\omega$ is  relative to  the unperturbed atomic line
in cm$^{-1}$, and comparable to the energy difference $\Delta V$ of
Fig. \ref{fig:potdiff} in the same units.}
\label{fig:NaKH2}
\end{figure}

\subsection{Study of the quasi-molecular satellites}
\label{sec:sat}

One outcome of our unified approach is that we may evaluate
the difference between the impact limit and the general unified profile, and
establish with certainty the region of validity of an assumed Lorentzian
profile.
Moreover the impact  approximation breaks down and  is not 
applicable even to the line core when an additional asymmetry 
is observed.

The unified profile shown in Fig.~\ref{fig:NaKH2lam} is the FT 
of the autocorrelation function given by Eq.~\ref{eq:phi}.
In radiative collision transitions it is the difference between the
final and initial state potential energies that determines the frequency and the energy of a single photon emitted
or absorbed by the interacting atoms.
The unified theory \citep{anderson1952,allard1978} predicts that there will
be line satellites centered periodically at 
frequencies corresponding to the extrema of the difference potential
between the upper and lower states.
In \citet{allard1980} we have studied the profiles of spectral lines of
alkalies perturbed by rare gases with a square-well potential, chosen
for its simplicity in order to understand the influence of potential
parameters on the profiles.
Blue satellite bands in the resonance lines of  Na/K-H$_2$ profiles
can be predicted from the  maxima in the difference potentials $\Delta V$
for the $B$-$X$ transition (Fig.~\ref{fig:potdiff}).
For clarity we only show  $\Delta V$ for  the symmetry C$_{2v}$.
Figure~8 of \cite{allard2007a}  shows  $\Delta V$ for other symmetries
that are all of the same order.
The difference potential maxima $\Delta V_{\rm max}$ are, respectively, 2500~cm$^{-1}$ for Na--H$_2$  and 1500~cm$^{-1}$ for K--H$_2$. The line satellite positions depend on the value of the extrema of the potential difference $\Delta V(R)$,
but they are always closer to the line center than this value. Line shape shows a strong density dependence: increasing the H$_2$ density results in a shift of the whole profile toward the position of the satellite feature (Fig.~\ref{fig:NaKH2}).
It is already the case  when the H$_2$  density reaches $1.2\times 10^{22}$~cm$^{-3}$ for the K resonance line whereas the corresponding Lorentzian profiles remains at the same position.
 In this illustration, $\sigma$ is a function of perturber density $n_p$, temperature $T$, and the frequency  difference (here in cm$^{-1}$), from the unperturbed spectral line center. 
 Figure~\ref{fig:NaKH2} shows that the line profiles of the Na/K resonance lines have orders of magnitude more absorption in the wing than a Lorentzian. The effect of this is increasingly important with increasing H$_2$ density.  As a result, the line profiles shift toward the position of the satellite band. There is a similar dependence on He density for the
Na/K resonance lines \citep{allard2023,allard2024}.
Lorentzian profiles used in \citet{siebenaler2025} to compute opacity tables until $10^5$~bar with large cutoffs are inadequate, since in that case Lorentzian profiles are not valid, even in the line core.

\subsection{Density range of validity  of a Lorentzian profile }
\label{sec:param}

Figure~\ref{fig:NaKH2lam}  clearly shows the asymmetrical shape of the very broad lines arising when the H$_2$ density reaches 10$^{22}$~cm$^{-3}$. While most apparent at this extreme, it  is already present at lower density for excited states.  Generally,  the higher the degree of excitation the more this is
the case, as has been illustrated by laboratory experiments that confirm the theory \citep{kielkopf1983}.
We conclude that the impact approximation breaks down and  is not 
applicable to calculate the line core 
when an additional asymmetry is observed due to the presence of a close line
satellite into the core of
the line profile,  even at low pressure, as in  Fig.~13 of
\citet{allard2016a} 
 for the $3p-4s$ transition of Mg-He
 and as in Fig.~6 of \citet{allard2012b}  for the
3$p$-4$s$ transitions of  Na perturbed by H$_2$.
Table~1 of the impact line parameters of Na--H$_2$  for the $3p-4s$
transition  of \citet{allard2012b}  and  of K--H$_2$ for the 
$4p-5s$ transition in \citet{deregt2025} are valid only at densities below
$10^{19}$~cm$^{-3}$ (0.3 bar).

\begin{figure}
 \centering
\resizebox{0.46\textwidth}{!}
{\includegraphics*{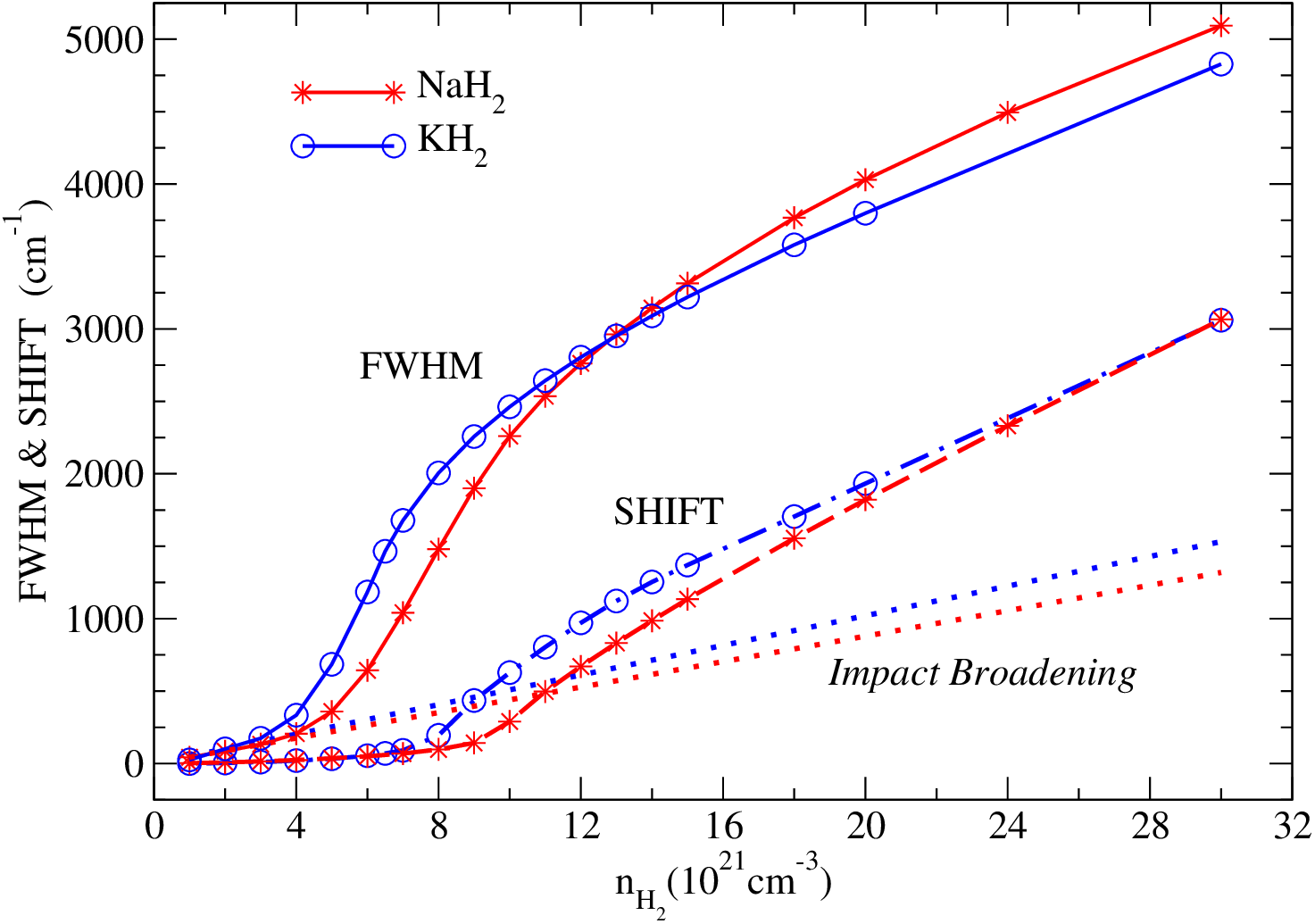}}
\caption  {The full width at half maximum (solid line) and shift (dashed line) of the Na (red) and K (blue) $D$2 spectral lines as a function of H$_2$ density.  The impact approximation FWHM (dotted line)  shown for comparison is invalid for all but the lowest densities.}
\label{fig:fwhmshift}
\end{figure}

\section{Conclusions}
The H$_2$ density range shown here probes the prominent quasi-molecular  features of the Na and K $D$2 components, respectively at 5889.95~\AA\/ and 7664.90~\AA\/. The impact approximation breaks down at a
density limit that is not necessarily very high.
The disagreement with a  Lorentzian profile depends on the value of
$\Delta V_{\mathrm{max}}$ in Fig.~\ref{fig:potdiff} responsible of the
appearance of quasi-molecular features.
The width, shift, and asymmetry
of the lines are then very dependent on the nature of
the interactions between the atoms at short and intermediate distance
when the perturber density is higher.

We see that the profiles and parameters of the  resonance lines 
 of Na/K perturbed by H$_2$ are thus very sensitive 
to H$_2$ density.   
For an H$_2$ density of $3 \times 10^{21}$ cm$^{-3}$,
the full width at half maximum (FWHM) of the resonance
lines of Na/K is larger than the impact approximation Lorentzian width  (Fig.~\ref{fig:fwhmshift}).
This is not a unique case, but one that clearly illustrates how 
non-linear the non-Lorentzian 
shapes can become outside the limited range of validity of the impact 
approximation usually
used in stellar and planetary atmosphere modeling.
We have investigated high H$_2$ broadened opacities of the doublets of the 
resonance lines of Na and K and shown that we have  to reject
Lorentzian approximations with unphysical values of a cut-off.
The use of the Lorentzian approximation with the line parameters of
\citet{allard2019,allard2016b} to calculate
opacity tables at high pressures up to $10^5$ bar for the Na/K--H$_2$ systems
in \citet{siebenaler2025} may lead to erroneous conclusions.

Data availability:
In \citet{allard2016b} only the red part of the line wing of the $D1$
    component was published online in CDS.\footnote{https://doi.org/10.26093/cds/vizier.35890021} 
  Complete K-H$_2$ opacity tables for the $D$1 and $D$2 components of the
       resonance line will be available at CDS anonymous FTP archive.

 On the other hand, online Ariel data in GitHub \citep{chubb2024} 
 contains a list of opacities and other data relevant for exoplanet atmospheres.\footnote{https://github.com/Ariel-data}
 The ExoMolOP opacity database  from \citep{chubb2021}
is on the
  ExoMol opacity website.\footnote{https://exomol.com/data/data-types/opacity}    
  
\begin{acknowledgements}
  We thank the referee for valuable comments that helped improve the
  manuscript. This work was supported by CNES, as part of the French
  contribution to the Ariel Space Mission.
  N.F. Allard is grateful to Gabriel-Dominique Marleau to point her the need
  to consider high density calculations of the Na/K resonance line profiles
  perturbed by H$_2$ and Pierre Drossart to useful discussions and
  encouragments to do this work.
\end{acknowledgements}

\bibliographystyle{aa} 

\begin{thebibliography}{48}
\expandafter\ifx\csname natexlab\endcsname\relax\def\natexlab#1{#1}\fi

\bibitem[{{Aglyamov} {et~al.}(2025){Aglyamov}, {Atreya}, {Bhattacharya}, {Li},
  {Levin}, {Bolton}, \& {Wong}}]{aglyamov2025}
{Aglyamov}, Y.~S., {Atreya}, S.~K., {Bhattacharya}, A., {et~al.} 2025, \icarus,
  425, 116334

\bibitem[{Allard {et~al.}(2007{\natexlab{a}})Allard, Allard, Homeier, Kielkopf,
  McCaughrean, \& Spiegelman}]{allard2007b}
Allard, F., Allard, N.~F., Homeier, D., {et~al.} 2007{\natexlab{a}}, A\&A, 474,
  L21

\bibitem[{Allard {et~al.}(2001)Allard, Hauschildt, Alexander, Tamanai, \&
  A.Schweitzer}]{allard2001}
Allard, F., Hauschildt, P.~H., Alexander, D.~R., Tamanai, A., \& A.Schweitzer.
  2001, Ap J., 556, 357

\bibitem[{Allard(1978)}]{allard1978}
Allard, N.~F. 1978, J. Phys. B: At. Mol. Opt. Phys., 11, 1383

\bibitem[{Allard {et~al.}(2003)Allard, Allard, Hauschildt, Kielkopf, \&
  Machin}]{allard2003}
Allard, N.~F., Allard, F., Hauschildt, P.~H., Kielkopf, J.~F., \& Machin, L.
  2003, A\&A, 411, L473

\bibitem[{Allard \& Biraud(1980)}]{allard1980}
Allard, N.~F. \& Biraud, Y.~G. 1980, J. Quant. Spectros. Radiat.Transfer., 23,
  253

\bibitem[{Allard \& Kielkopf(1982)}]{allard1982}
Allard, N.~F. \& Kielkopf, J.~F. 1982, Rev. Mod. Phys., 54, 1103

\bibitem[{{Allard} {et~al.}(2024){Allard}, {Kielkopf}, {Myneni}, \&
  {Blakely}}]{allard2024}
{Allard}, N.~F., {Kielkopf}, J.~F., {Myneni}, K., \& {Blakely}, J.~N. 2024,
  \aap, 683, A188

\bibitem[{Allard \& Koester(1992)}]{allard1992}
Allard, N.~F. \& Koester, D. 1992, A\&A, 258, 464

\bibitem[{Allard {et~al.}(1994)Allard, Koester, Feautrier, \&
  Spielfiedel}]{allard1994}
Allard, N.~F., Koester, D., Feautrier, N., \& Spielfiedel, A. 1994, A\&A
  Suppl., 108, 417

\bibitem[{Allard {et~al.}(2016a)Allard, Leininger, Gad\'ea, Brousseau-Couture,
  \& Dufour}]{allard2016a}
Allard, N.~F., Leininger, T., Gad\'ea, F.~X., Brousseau-Couture, V., \& Dufour,
  P. 2016a, A\&A, 588, A142

\bibitem[{{Allard} {et~al.}(2023){Allard}, {Myneni}, {Blakely}, \&
  {Guillon}}]{allard2023}
{Allard}, N.~F., {Myneni}, K., {Blakely}, J.~N., \& {Guillon}, G. 2023, \aap,
  674, A171

\bibitem[{Allard {et~al.}(1999)Allard, Royer, Kielkopf, \&
  Feautrier}]{allard1999}
Allard, N.~F., Royer, A., Kielkopf, J.~F., \& Feautrier, N. 1999, Phys. Rev. A,
  60, 1021

\bibitem[{Allard {et~al.}(2007{\natexlab{b}})Allard, Spiegelman, \&
  Kielkopf}]{allard2007a}
Allard, N.~F., Spiegelman, F., \& Kielkopf, J.~F. 2007{\natexlab{b}}, A\&A,
  465, 1085

\bibitem[{{Allard} {et~al.}(2016b){Allard}, {Spiegelman}, \&
  {Kielkopf}}]{allard2016b}
{Allard}, N.~F., {Spiegelman}, F., \& {Kielkopf}, J.~F. 2016b, A\&A, 589, A21

\bibitem[{Allard {et~al.}(2012)Allard, Spiegelman, Kielkopf, Tinetti, \&
  Beaulieu}]{allard2012b}
Allard, N.~F., Spiegelman, F., Kielkopf, J.~F., Tinetti, G., \& Beaulieu, J.~P.
  2012, A\&A, 543, A159

\bibitem[{{Allard} {et~al.}(2019){Allard}, {Spiegelman}, {Leininger}, \&
  {Molliere}}]{allard2019}
{Allard}, N.~F., {Spiegelman}, F., {Leininger}, T., \& {Molliere}, P. 2019,
  \aap, 628, A120

\bibitem[{{Anderson}(1952)}]{anderson1952}
{Anderson}, P.~W. 1952, Physical Review, 86, 809

\bibitem[{Baranger(1958{\natexlab{a}})}]{baranger1958c}
Baranger, M. 1958{\natexlab{a}}, Phys. Rev., 112, 855

\bibitem[{Baranger(1958{\natexlab{b}})}]{baranger1958b}
Baranger, M. 1958{\natexlab{b}}, Phys. Rev., 111, 494

\bibitem[{Baranger(1958{\natexlab{c}})}]{baranger1958a}
Baranger, M. 1958{\natexlab{c}}, Phys. Rev., 111, 481

\bibitem[{{Blouin} {et~al.}(2019){Blouin}, {Dufour}, {Allard}, {Salim}, {Rich},
  \& {Koopmans}}]{blouin2019c}
{Blouin}, S., {Dufour}, P., {Allard}, N.~F., {et~al.} 2019, \apj, 872, 188

\bibitem[{Ch'en \& Takeo(1957)}]{chen1957}
Ch'en, S.~Y. \& Takeo, M. 1957, Rev. Mod. Phys., 29, 20

\bibitem[{{Chubb} {et~al.}(2021){Chubb}, {Rocchetto}, {Yurchenko}, {Min},
  {Waldmann}, {Barstow}, {Molli{\`e}re}, {Al-Refaie}, {Phillips}, \&
  {Tennyson}}]{chubb2021}
{Chubb}, K.~L., {Rocchetto}, M., {Yurchenko}, S.~N., {et~al.} 2021, \aap, 646,
  A21

\bibitem[{{Chubb} {et~al.}(2024){Chubb}, {Robert}, {Sousa-Silva}, {Yurchenko},
  {Allard}, {Boudon}, {Buldyreva}, {Bultel}, {Coustenis}, {Foltynowicz},
  {Gordon}, {Hargreaves}, {Helling}, {Hill}, {Hrodmarsson}, {Karman},
  {Lecoq-Molinos}, {Migliorini}, {Rey}, {Richard}, {Sadiek}, {Schmidt},
  {Sokolov}, {Stefani}, {Tennyson}, {Venot}, {Wright}, {Arenales-Lope},
  {Barstow}, {Bocchieri}, {Carrasco}, {Dubey}, {Egorov}, {Mu{\~n}oz},
  {Gharib-Nezhad}, {Gkouvelis}, {Gr{\"u}bel}, {Irwin}, {Kn{\'\i}{\v{z}}ek},
  {Lewis}, {Lodge}, {Ma}, {Martins}, {Molaverdikhani}, {Morello}, {Nikitin},
  {Panek}, {Rengel}, {Rinaldi}, {Skinner}, {Tinetti}, {van Kempen}, {Yang}, \&
  {Zingales}}]{chubb2024}
{Chubb}, K.~L., {Robert}, S., {Sousa-Silva}, C., {et~al.} 2024, RAS Techniques
  and Instruments, 3, 636

\bibitem[{CODATA(2025)}]{codata2025}
CODATA, N. 2025, Fundamental Physical Constants,
  \url{https://physics.nist.gov/cgi-bin/cuu/Value?n0}, accessed: 2025-05-31

\bibitem[{{de Regt} {et~al.}(2025){de Regt}, {Snellen}, {Allard}, {Gonz{\'a}lez
  Picos}, {Gandhi}, {Grasser}, {Landman}, {Molli{\`e}re}, {Nasedkin},
  {Stolker}, \& {Zhang}}]{deregt2025}
{de Regt}, S., {Snellen}, I.~A.~G., {Allard}, N.~F., {et~al.} 2025, \aap, 696,
  A225

\bibitem[{Fano(1963)}]{fano1963}
Fano, F. 1963, Phys. Rev., 131, 259

\bibitem[{{Guillot} {et~al.}(1994){Guillot}, {Gautier}, {Chabrier}, \&
  {Mosser}}]{guillot1994}
{Guillot}, T., {Gautier}, D., {Chabrier}, G., \& {Mosser}, B. 1994, \icarus,
  112, 337

\bibitem[{Herman \& Sando(1978)}]{herman1978}
Herman, P.~S. \& Sando, K.~M. 1978, J. Chem. Phys., 68, 1152

\bibitem[{{Homeier} {et~al.}(2007){Homeier}, {Allard}, {Johnas}, {Hauschildt},
  \& {Allard}}]{homeier2007}
{Homeier}, D., {Allard}, N., {Johnas}, C.~M.~S., {Hauschildt}, P.~H., \&
  {Allard}, F. 2007, in Astronomical Society of the Pacific Conference Series,
  Vol. 372, 15th European Workshop on White Dwarfs, ed. {R.~Napiwotzki \&
  M.~R.~Burleigh}, 277

\bibitem[{{Hubeny} \& {Mihalas}(2015)}]{hubeny2015}
{Hubeny}, I. \& {Mihalas}, D. 2015, {Theory of Stellar Atmospheres. An
  Introduction to Astrophysical Non-equilibrium Quantitative Spectroscopic
  Analysis}, Princeton Series in Astrophysics (Princeton and Oxford: Princeton
  University Press)

\bibitem[{Jablonski(1945)}]{jablonski1945}
Jablonski, A. 1945, Phys. Rev., 68, 78

\bibitem[{Jablonski(1948)}]{jablonski1948}
Jablonski, A. 1948, Phys. Rev., 73, 258

\bibitem[{{Janssen} {et~al.}(2017){Janssen}, {Oswald}, {Brown}, {Gulkis},
  {Levin}, {Bolton}, {Allison}, {Atreya}, {Gautier}, {Ingersoll}, {Lunine},
  {Orton}, {Owen}, {Steffes}, {Adumitroaie}, {Bellotti}, {Jewell}, {Li}, {Li},
  {Misra}, {Oyafuso}, {Santos-Costa}, {Sarkissian}, {Williamson}, {Arballo},
  {Kitiyakara}, {Ulloa-Severino}, {Chen}, {Maiwald}, {Sahakian}, {Pingree},
  {Lee}, {Mazer}, {Redick}, {Hodges}, {Hughes}, {Bedrosian}, {Dawson}, {Hatch},
  {Russell}, {Chamberlain}, {Zawadski}, {Khayatian}, {Franklin}, {Conley},
  {Kempenaar}, {Loo}, {Sunada}, {Vorperion}, \& {Wang}}]{janssen2017}
{Janssen}, M.~A., {Oswald}, J.~E., {Brown}, S.~T., {et~al.} 2017, \ssr, 213,
  139

\bibitem[{Julienne \& Mies(1986)}]{julienne1986}
Julienne, P.~S. \& Mies, F.~H. 1986, Phys. Rev. A, 34, 3792

\bibitem[{{Kielkopf}(1983)}]{kielkopf1983}
{Kielkopf}, J. 1983, Journal of Physics B Atomic Molecular Physics, 16, 3149

\bibitem[{{Kielkopf} \& {Allard}(1979)}]{kielkopf1979}
{Kielkopf}, J.~F. \& {Allard}, N.~F. 1979, Physical Review Letters, 43, 196

\bibitem[{Kielkopf {et~al.}(2017)Kielkopf, Allard, Alekseev, Spiegelman,
  Guillon, \& Berriche}]{kielkopf2017}
Kielkopf, J.~F., Allard, N.~F., Alekseev, V.~A., {et~al.} 2017, Journal of
  Physics: Conference Series, 810, 012023

\bibitem[{Margenau \& Watson(1936)}]{margenau1936}
Margenau, H. \& Watson, W. 1936, Rev. Mod. Phys., 8, 22

\bibitem[{Mies {et~al.}(1986)Mies, Julienne, Band, \& Singer}]{mies1986}
Mies, F.~H., Julienne, P.~S., Band, Y.~B., \& Singer, S.~J. 1986, J. Phys. B:
  At. Mol. Opt. Phys., 19, 3249

\bibitem[{{Militzer} \& {Hubbard}(2024)}]{militzer2024}
{Militzer}, B. \& {Hubbard}, W.~B. 2024, \icarus, 411, 115955

\bibitem[{Royer(1971)}]{royer1971a}
Royer, A. 1971, Phys. Rev. A, 3, 2044

\bibitem[{Royer(1974)}]{royer1974}
Royer, A. 1974, Can. J. Phys., 52, 1816

\bibitem[{Royer(1980)}]{royer1980}
Royer, A. 1980, Phys. Rev. A, 22, 1625

\bibitem[{Sando \& Herman(1983)}]{sando1983}
Sando, K.~M. \& Herman, P.~S. 1983, in Spectral Line Shapes, Vol.~2 (New York:
  Walter de Gruyter), 497

\bibitem[{{Siebenaler} {et~al.}(2025){Siebenaler}, {Miguel}, {de Regt}, \&
  {Guillot}}]{siebenaler2025}
{Siebenaler}, L., {Miguel}, Y., {de Regt}, S., \& {Guillot}, T. 2025, \aap,
  693, A308

\bibitem[{{Stevenson}(1982)}]{stevenson1982}
{Stevenson}, D.~J. 1982, Annual Review of Earth and Planetary Sciences, 10, 257

\bibitem[{Szudy \& Baylis(1975)}]{szudy1975}
Szudy, J. \& Baylis, W. 1975, J. Quant. Spectrosc. Radiat. Transfer, 15, 641

\bibitem[{Tinetti {et~al.}(2007)Tinetti, Vidal-Madjar, Liang, Beaulieu, Yung,
  Carey, Barber, Tennyson, Ribas, Allard, Ballester, Sing, \&
  Selsis}]{tinetti2007}
Tinetti, G., Vidal-Madjar, A., Liang, M.-C., {et~al.} 2007, Nature, 448, 169

\end{thebibliography}

\end{document}